\documentclass[conference]{IEEEtran}
\usepackage[top = 0.76 in, bottom = 1.04 in, left = 0.62in, right = 0.62 in]{geometry}
\IEEEoverridecommandlockouts
\usepackage{cite}
\usepackage{amsmath,amssymb,amsfonts}
\usepackage{algorithmic}
\usepackage{graphicx}
\usepackage{textcomp}
\usepackage{multirow}
\usepackage{xcolor}
\def\BibTeX{{\rm B\kern-.05em{\sc i\kern-.025em b}\kern-.08em
    T\kern-.1667em\lower.7ex\hbox{E}\kern-.125emX}}
\begin{document}

\title{Path Loss Modeling for NLoS Ultraviolet Channels Incorporating Scattering and Reflection Effects}

\author{\IEEEauthorblockN{Tianfeng Wu\textsuperscript{\dag}, Fang Yang\textsuperscript{\dag}, Fei Li\textsuperscript{$\natural$}, Renzhi Yuan\textsuperscript{\ddag}, Tian Cao\textsuperscript{\S}, Ling Cheng\textsuperscript{$\ast$},\\Jian Song\textsuperscript{\dag,}\textsuperscript{$\sharp$}, Julian Cheng\textsuperscript{$\star$}, and Zhu Han\textsuperscript{$\uplus$}}\\ 

\IEEEauthorblockA{\textsuperscript{\dag}Department of Electronic Engineering, BNRist, Tsinghua University, Beijing 100084, P. R. China \\ \textsuperscript{$\natural$}Hanjiang National Laboratory, Wuhan 430060, P. R. China\\ \textsuperscript{\ddag}State Key Laboratory of Networking and Switching Technology,\\Beijing University of Posts and Telecommunications, Beijing, 100876, P. R. China\\ \textsuperscript{\S}School of Telecommunications Engineering, Xidian University, Xi’an 710071, P. R. China\\ \textsuperscript{$\ast$}School of Electrical and Information Engineering, University of the Witwatersrand, Johannesburg 2000, South Africa\\ \textsuperscript{$\sharp$}Shenzhen International Graduate School, Tsinghua University, Shenzhen 518055, P. R. China\\ \textsuperscript{$\star$}School of Engineering, The University of British Columbia, Kelowna, BC, V1V 1V7, Canada\\ \textsuperscript{$\uplus$}Department of Electrical and Computer Engineering, University of Houston, Houston, TX 77004 USA\\Email: wtf22@mails.tsinghua.edu.cn, fangyang@tsinghua.edu.cn, buaafeili@126.com, renzhi.yuan@bupt.edu.cn,\\caotian@xidian.edu.cn, ling.cheng@wits.ac.za, jsong@tsinghua.edu.cn, julian.cheng@ubc.ca, hanzhu22@gmail.com}}

\maketitle

\begin{abstract}
This paper tackles limitations in existing non-line-of-sight (NLoS) ultraviolet (UV) channel models, where conventional approaches assume obstacle-free propagation or uniform radiation intensity. In this paper, we develop a path loss model incorporating scattering and reflection, and then propose an obstacle-boundary approximation method to achieve computational tractability. Our framework systematically incorporates spatial obstacle properties, including dimensions, coordinates, contours, and orientation angles, while employing the Lambertian radiation pattern for source modeling. Additionally, the proposed path loss model is validated by comparing it with the Monte-Carlo photon-tracing model and analytical integral model via numerical results, which indicate that when obstacle reflection is prominent, an approximation treatment of obstacle boundaries has a negligible influence on the path loss estimation of NLoS UV communication channels.           
\end{abstract}

\begin{IEEEkeywords}
Path loss model, scattering, reflection, NLoS UV communication channels, Lambertian distribution.   
\end{IEEEkeywords}

\section{Introduction}
\IEEEPARstart{R}{adio} frequency technology is widely utilized because of low cost and flexible deployment, yet struggles with spectral congestion, security risks, and interference issues \cite{ref1}. While alternatives like millimeter-wave and visible light communications improve specific aspects, their line-of-sight requirements and obstruction sensitivity restrain practicality \cite{ref2}. Recent advances in ultraviolet (UV) communications (200-280 nm solar-blind band) offer transformative potential by leveraging unique atmospheric characteristics \cite{ref3}. UV signals achieve license-free operation with minimal solar noise through ozone absorption while enabling non-line-of-sight (NLoS) links via atmospheric scattering \cite{ref4}. The technology inherently resists eavesdropping and electromagnetic interference \cite{ref5}, making it ideal for secure dense networks in complex environments \cite{ref6}. By integrating with existing wireless technologies, UV communications could redefine reliable connectivity frameworks for next-generation communication systems through its advantages.

Research in UV communications has prioritized NLoS channel modeling due to its fundamental impact on system design optimization \cite{ref7}, where the scattering order of UV photons serves as the key physical determinant. Specifically, single-scattering models [8]--[10] prove effective for characterizing short-range systems, focusing on essential parameters like the angular distribution of scattered photons and temporal broadening effects in impulse response. These metrics directly affect signal detection accuracy in proximity-based applications. Conversely, long-range UV systems require multi-scattering models [11]--[13] to address complex propagation phenomena, particularly bandwidth constraints caused by delayed signal components and the accumulated path loss variations arising from successive scattering interactions. Recent progress further enhances NLoS modeling through obstruction-aware frameworks \cite{ref7,ref14,ref15}, which provide systematic methods to quantify how physical barriers alter signal propagation patterns and energy distribution.

Although these works lay a solid theoretical foundation for UV channel modeling, they also encounter certain challenges with the ongoing complexity of communication environments. Specifically, the existing research \cite{ref7,ref15} considers the obstacle shape as a cuboid, while in practical scenarios, the shape of the obstacle can be diverse. Moreover, when solving for the received energy contributed by air scattering, the methodology adopted in \cite{ref7} to ascertain whether the propagation links of UV photons are blocked by the obstacle is comparatively complex, including many scenarios, cases, and conditions caused by the intersection between transceiver field-of-views (FoVs) and obstacle, which brings certain difficulties to practical applications. 

To address the problems mentioned above, we propose a path loss model incorporating scattering and reflection effects in this paper, where an obstacle-boundary approximation method is put forward to reduce modeling complexity. Moreover, the spatial properties of the obstacle, including dimensions, coordinates, contours, and orientation angles, are considered to emulate practical communication environments. Furthermore, the proposed path loss model is validated by comparing it with the Monte-Carlo photon-tracing (MCPT) model and integral model through numerical results.   

\textit{Notations}: Throughout this paper, boldface lowercase letters (e.g. $\boldsymbol{n}$) and boldface uppercase letter pairs (e.g. $\boldsymbol{\rm{RS}}$) represent vectors, and $||\cdot||$ denotes the Euclidean norm of a vector.   

\begin{figure}  
\centering  
\includegraphics[scale=0.42]{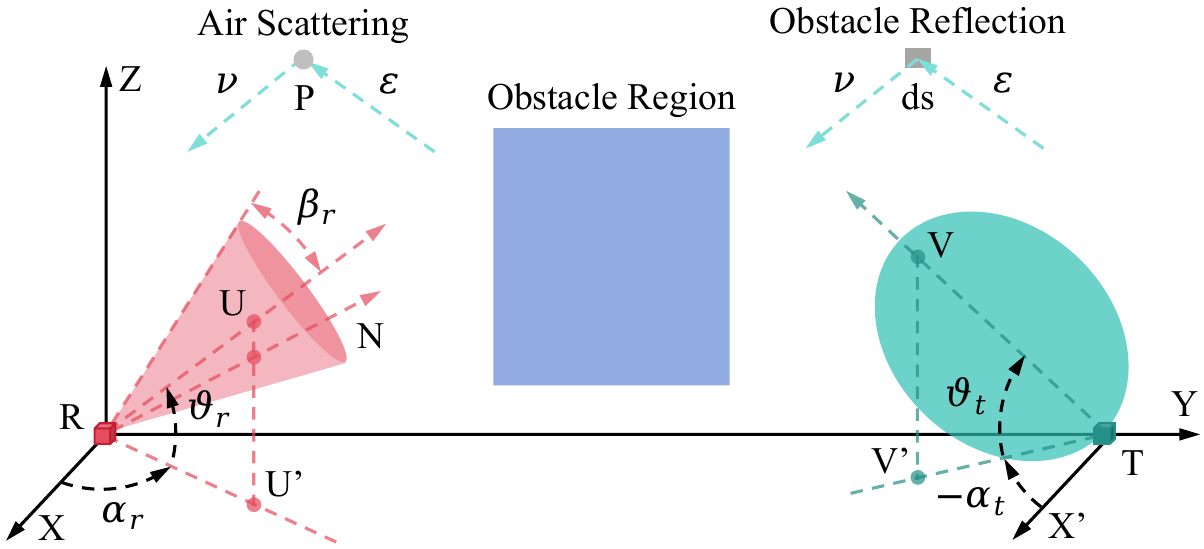}
\centering
\caption{Diagram of NLoS UV communication scenarios considering obstacles.}
\label{Fig1}  
\end{figure} 
\begin{figure}  
\centering  
\includegraphics[scale=0.56]{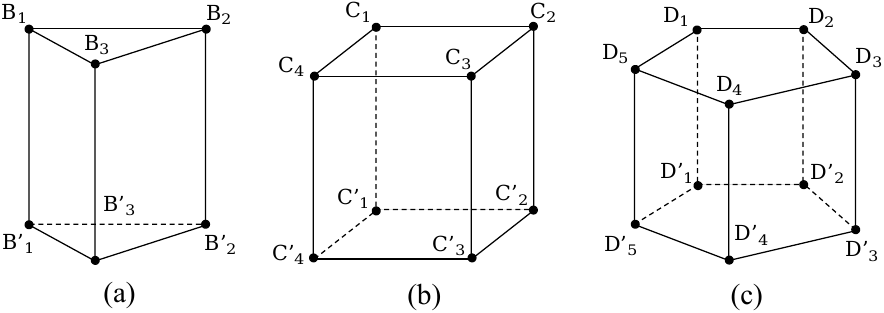}
\centering
\caption{Obstacle shapes in the obstacle region: (a) RTP, (b) RP, and (c) RPP.}
\label{Fig2}  
\end{figure} 

\section{System Model and Parameter Definitions}  
Combined with the actual communication environments, the diagram of NLoS communication scenarios is shown in Fig.~\ref{Fig1}, where a light-emitting diode (LED) is employed as the UV light source. We consider three representative obstacle shapes in the obstacle region, namely the regular triangular prism (RTP), the rectangular prism (RP), and the regular pentagonal prism (RPP), as depicted in Fig.~\ref{Fig2}.

\begin{figure}  
\centering  
\includegraphics[scale=0.54]{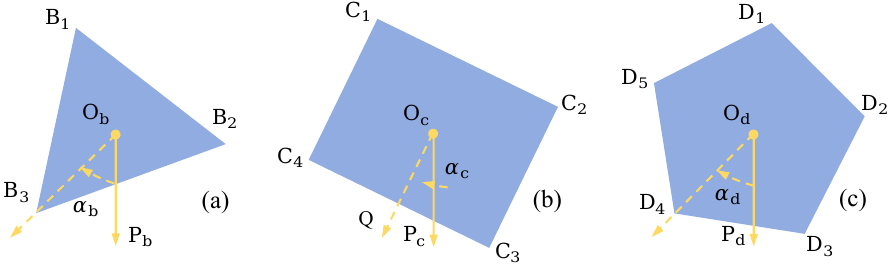}
\centering
\caption{Illustration of orientation angles for different obstacle shapes: (a) RTP, (b) RP, and (c) RPP.}
\label{Fig3}  
\end{figure}

The parameters of the system model are defined as follows: $\beta_r$ denotes the half FoV angle of the receiver; $\vartheta_r$ denotes the receiver (R) elevation angle and is positive if taken anticlockwise from RU', where RU' is the projection of the FoV axis RU on the plane XRY; $\vartheta_t$ denotes the transmitter (T) elevation angle and is positive if taken clockwise from the projection of the beam axis TV on the plane XRY; $\alpha_r$ represents the receiver azimuth angle, which is positive if rotating counterclockwise from the X positive axis; $\alpha_t$ denotes the transmitter azimuth angle and is positive if rotating counterclockwise from $\boldsymbol{\rm{TX'}}$; $r$ represents the communication range; $\varepsilon$ and $\nu$ are the distances from the scattering point P (or the differential reflection area d$s$) to T and R, respectively; and $\mathcal{A}$ represents the detection area of the receiver aperture. Besides, $\alpha_{b}$ denotes the orientation angle of RTP, which is positive if rotating clockwise from $\boldsymbol{{\rm{O}}_b {\rm{P}}_b}$ to $\boldsymbol{{\rm{O}}_b {\rm{B}}_3}$, as shown in Fig.~\ref{Fig3}; $\alpha_c$ represents the orientation angle of RP and is positive if taken clockwise from $\boldsymbol{{\rm{O}}_c {\rm{P}}_c}$ to $\boldsymbol{{\rm{O}}_c \rm{Q}}$, where $\boldsymbol{{\rm{O}}_c \rm{Q}}$ is perpendicular to the line $\rm{C_3 C_4}$; $\alpha_d$ is the orientation angle of RPP, which is positive if rotating clockwise from $\boldsymbol{{\rm{O}}_d {\rm{P}}_d}$ to $\boldsymbol{{\rm{O}}_d {\rm{D}}_4}$; ${\rm{O}}_b$, ${\rm{O}}_c$, and ${\rm{O}}_d$ are the central points of RTP, RP, and RPP, respectively; $\omega_b$, $\omega_c$, $\mu_c$, and $\omega_d$ are the lengths of $\rm{B_1 B_2}$, $\rm{C_1 C_2}$, $\rm{C_2 C_3}$, and $\rm{D_1 D_2}$, respectively; $\gamma_b$, $\gamma_c$, and $\gamma_d$ denote the heights of RTP, RP, and RPP, respectively.

Next, we specify the intervals of several variables mentioned above. Considering the symmetries of the obstacles, we set the intervals of $\alpha_b$, $\alpha_c$, and $\alpha_d$ to $[0,2\pi/3]$, $[0,\pi]$, and $[0,2\pi/5]$, respectively, and set the active areas of the points ${\rm{O}}_b(x_b,y_b,z_b)$, ${\rm{O}}_c(x_c,y_c,z_c)$, and ${\rm{O}}_d(x_d,y_d,z_d)$ to
\begin{equation}
{\rm{O_{\xi}}}:
\begin{cases}
x_{\xi}\in(-\infty,-\Gamma^{sha}_{\rm{rad}}),\\
y_{\xi}\in(\Gamma^{sha}_{\rm{rad}},r-\Gamma^{sha}_{\rm{rad}}),\\
z_{\xi}=\gamma_{\xi},
\end{cases}
\end{equation} 
where $\xi$ can be $b$, $c$, and $d$, ``$sha$'' can be ``RTP'', ``RP'', and ``RPP'', and $\Gamma^{sha}_{\rm{rad}}$ is the rotation radius of the obstacle. Based on the geometric relationships, $\Gamma^{sha}_{\rm{rad}}$ can be expressed as    
\begin{equation}  
\Gamma^{sha}_{\rm{rad}}=
\begin{cases}
{\omega_b}/{\sqrt{3}}, &\rm{RTP},\\
{\sqrt{\omega_c^2+\mu_c^2}}/{2}, &\rm{RP},\\ 
{\omega_d\sin(3\pi/10)}/{\sin(2\pi/5)}, &\rm{RPP},\\
\end{cases}               
\end{equation} 
Therefore, the coordinates of \{$\rm{B_1}$, $\rm{B_2}$, $\rm{B_3}$\}, \{$\rm{C_1}$, $\rm{C_2}$, $\rm{C_3}$, $\rm{C_4}$\}, and \{$\rm{D_1}$, $\rm{D_2}$, $\rm{D_3}$, $\rm{D_4}$, $\rm{D_5}$\} can be derived as
\begin{subequations}
\begin{align}
&
\begin{cases}
x_{\xi_b}=-\Gamma^{\rm{RTP}}_{\rm{rad}}\sin(\alpha_b+\varpi_b)+x_b,\\
y_{\xi_b}=-\Gamma^{\rm{RTP}}_{\rm{rad}}\cos(\alpha_b+\varpi_b)+y_b,\\
z_{\xi_b}=\gamma_b,
\end{cases}
\\&
\begin{cases}
x_{\xi_c}=-\Gamma^{\rm{RP}}_{\rm{rad}}\sin(\alpha_c+\varpi_c)+x_c,\\
y_{\xi_c}=-\Gamma^{\rm{RP}}_{\rm{rad}}\cos(\alpha_c+\varpi_c)+y_c,\\
z_{\xi_c}=\gamma_c,
\end{cases}
\\&
\begin{cases}
x_{\xi_d}=-\Gamma^{\rm{RPP}}_{\rm{rad}}\sin(\alpha_d+\varpi_d)+x_d,\\
y_{\xi_d}=-\Gamma^{\rm{RPP}}_{\rm{rad}}\cos(\alpha_d+\varpi_d)+y_d,\\
z_{\xi_d}=\gamma_d,
\end{cases}
\end{align}
\end{subequations} 
respectively, where $\xi_b$ can be $\{b_1, b_2, b_3\}$, and the values of $\varpi_b$ for $\rm{B_1}$, $\rm{B_2}$, and $\rm{B_3}$ are $\pi/6$, $5\pi/6$, and $3\pi/2$, respectively; $\xi_c$ can be $\{c_1, c_2, c_3, c_4\}$, and the values of $\varpi_c$ for $\rm{C_1}$, $\rm{C_2}$, $\rm{C_3}$, and $\rm{C_4}$ are $\tan^{-1}(\mu_c/\omega_c)$, $\pi-\tan^{-1}(\mu_c/\omega_c)$, $\pi+\tan^{-1}(\mu_c/\omega_c)$, and $2\pi-\tan^{-1}(\mu_c/\omega_c)$, respectively; $\xi_d$ can be $\{d_1, d_2, d_3, d_4, d_5\}$, and the values of $\varpi_d$ for $\rm{D_1}$, $\rm{D_2}$, $\rm{D_3}$, $\rm{D_4}$, and $\rm{D_5}$ are $3\pi/10$, $7\pi/10$, $11\pi/10$, $3\pi/2$, and $19\pi/10$, respectively. 

Besides, we specify the intervals of the transceiver elevation angles, azimuth angles, and the receiver FoV angle. Combining with the active areas of the obstacles, $\beta_r$, $\vartheta_t$, and $\vartheta_r$ are all set to $(0,\pi/2)$, and $\alpha_t$ and $\alpha_r$ are set to $(-\pi,-\pi/2)$ and $(\pi/2,$ $\pi)$, respectively, which satisfy the commands of general NLoS UV communication scenarios.  

When a pulse of energy $\mathcal{Q}_t$ is transmitted by the transmitter, the received pulse energy is composed of two parts: the energy contributed by air scattering, $\mathcal{Q}_{r,\rm{sca}}$, and the energy contributed by obstacle reflection, $\mathcal{Q}_{r,\rm{ref}}$. Thus, the path loss of NLoS UV communication channels can be defined as
\begin{equation}
\mathcal{L}[{\rm{dB}}]=10\log_{10}[1/(\mathcal{Q}_{r,\rm{sca}}+\mathcal{Q}_{r,\rm{ref}})].                        
\end{equation}
In the following, we will derive $\mathcal{Q}_{r,\rm{sca}}$ and $\mathcal{Q}_{r,\rm{ref}}$ in conjunction with specific obstacle contours.

\section{Scattered Energy Derivation} 
Referring to the existing single-scattering propagation theory \cite{ref7,ref14}, the received pulse energy obtained from air scattering can be derived as
\begin{equation}
\mathcal{Q}_{r,\rm{sca}}=\int_{\vartheta_{\min}}^{\vartheta_{\max}}\int_{\psi_{\min}}^{\psi_{\max}}\int_{\nu_{\min}}^{\nu_{\max}}\mathcal{Q}_t k_s \mathcal{A}\,\mathcal{G}\,\mathcal{S}_{\rm{wei}}\,{\rm{d}}\nu{\rm{d}}\psi{\rm{d}}\vartheta,
\label{e4}
\end{equation} 
where
\begin{equation}
\mathcal{G}=\displaystyle{\frac{\kappa+1}{2\pi\varepsilon^2}}\cos^{\kappa}\varpi {\rm{P}}(\cos{\beta}) \cos{\delta} \exp[-k_e(\varepsilon+\nu)] \cos\psi. 
\end{equation} 
Here, $\varpi$ denotes the angle between the transmitter pointing direction and photon emitting direction. $\kappa$ is the order of Lambertian emission and can be given by $-\ln2/\ln\cos(\beta_{1/2}/2)$, where $\beta_{1/2}$ denotes the full-width at half-illuminance of the LED. $\psi$ denotes the angle between \textbf{RN} and \textbf{RS} (\textbf{RS} is located on the plane $\mathcal{E}_{\vartheta}$ and enclosed by the receiver FoV), which is positive when rotating anticlockwise from \textbf{RN} as shown in Fig.~\ref{Fig1}. $\mathcal{E}_{\vartheta}$ represents the plane passing through the ray RN and is perpendicular to the plane RUU', which rotates around the line RW, where RW is located on the XRY and perpendicular to the line RU', and the subscript $\vartheta$ denotes the angle between \textbf{RU} and \textbf{RN}, which is positive if taken anticlockwise from \textbf{RU}. $k_e$ is the extinction coefficient of the atmosphere, which can be obtained by adding the scattering coefficient $k_s$ and the absorption coefficient $k_a$. $\beta$ and $\delta$ denote the angles between \textbf{TP} and \textbf{PR}, and \textbf{RU} and \textbf{RP}, respectively, $\mathcal{S}_{\rm{wei}}$ is the weighting factor, and ${\rm{P}}(\cos{\beta})$ is the scattering phase function \cite{ref6}. 

The unknown quantities in (\ref{e4}) are $\vartheta_{\min}$, $\vartheta_{\max}$, $\psi_{\min}$, $\psi_{\max}$, $\nu_{\min}$, $\nu_{\max}$, and $\mathcal{S}_{\rm{wei}}$. Given the wide coverage area of the light source based on Lambertian distribution, the upper and lower bounds of the triple integral in $\mathcal{Q}_{r,\rm{sca}}$ are mainly determined by the receiver conical surface, and can be derived as
\begin{equation}
\vartheta_{\min}=-\beta_r,\,\,\vartheta_{\max}=\beta_r, 
\label{e6}
\end{equation}             
\begin{equation}
\psi_{\max}=\tan^{-1}\sqrt{\cos^2{\vartheta}\tan^2{\beta_r}-\sin^2{\vartheta}}=-\psi_{\min},
\label{e7}
\end{equation} 
\begin{equation}
\nu_{\min}=0,\,\,\nu_{\max}=+\infty, 
\label{e8}
\end{equation} 
respectively. Meanwhile, a constraint needs to be imposed on the scattering point P, which can be expressed as
\begin{equation}
\frac{1}{\varepsilon}\,{\boldsymbol{\mu_t}}{\boldsymbol{\varepsilon}}^{\rm{T}}\geq\cos\frac{\pi}{2},\,\,\,{\rm{i.e.}},\,\,\,{\boldsymbol{\mu_t}}{\boldsymbol{\varepsilon}}^{\rm{T}}\geq0,
\label{e9}
\end{equation}
where ${\boldsymbol{\mu_t}}=[\cos\vartheta_t\cos\alpha_t,\cos\vartheta_t\sin\alpha_t,\sin\vartheta_t]$ and ${\boldsymbol{\varepsilon}}=[x,$ $y-r, z]$. Moreover, the coordinates of P can be given by 
\begin{equation}
{\rm{P}:}
\begin{cases}
x=\nu\cos{\psi}\cos{\phi}\,{\cos(\alpha_r+\omega)}{\sec{\omega}},\\
y=\nu\cos{\psi}\cos{\phi}\,{\sin(\alpha_r+\omega)}{\sec{\omega}},\\
z=\nu\cos{\psi}\sin{\phi}.
\end{cases}
\label{e10}
\end{equation} 
Here, $\phi=\vartheta_r+\vartheta$ and $\omega=\tan^{-1}(\tan\psi/\cos\phi)$. $\phi_{\min}=\vartheta_r+\vartheta_{\min}$ is assumed to be larger than $-\pi/2$, and $\phi_{\max}=\vartheta_r+\vartheta_{\max}$ is assumed to be less than $\pi/2$. In the following, we derive the values of $\mathcal{S}_{\rm{wei}}$ corresponding to subfigures of Fig.~\ref{Fig2} under the condition that the inequality (\ref{e9}) holds. 

Initially, we propose an obstacle-boundary approximation method to address the boundaries of RTP, RP, and RPP, where the critical parameters $\Upsilon^{sha}_{r,\rm{upp}}$, $\Upsilon^{sha}_{r,\min}$, $\Upsilon^{sha}_{r,\max}$, $\Upsilon^{sha}_{t,\rm{upp}}$, $\Upsilon^{sha}_{t,\min}$, and $\Upsilon^{sha}_{t,\max}$ are introduced for tractable analysis. Based on the geometric relationships, $\Upsilon^{sha}_{r,\rm{upp}}$ and $\Upsilon^{sha}_{t,\rm{upp}}$ can be derived as
\begin{align}
\Upsilon^{sha}_{r,\rm{upp}}=&
\begin{cases}
\max(\Upsilon^{\rm{RTP}}_{r,b_1},\!\Upsilon^{\rm{RTP}}_{r,b_2},\!\Upsilon^{\rm{RTP}}_{r,b_3}),\\
\max(\Upsilon^{\rm{RP}}_{r,c_1},\!\Upsilon^{\rm{RP}}_{r,c_2},\!\Upsilon^{\rm{RP}}_{r,c_3},\!\Upsilon^{\rm{RP}}_{r,c_4}),\\
\max(\Upsilon^{\rm{RPP}}_{r,d_1},\!\Upsilon^{\rm{RPP}}_{r,d_2},\!\Upsilon^{\rm{RPP}}_{r,d_3},\!\Upsilon^{\rm{RPP}}_{r,d_4},\!\Upsilon^{\rm{RPP}}_{r,d_5}),\\
\end{cases}\\
\Upsilon^{sha}_{t,\rm{upp}}=&
\begin{cases}
\max(\Upsilon^{\rm{RTP}}_{t,b_1},\Upsilon^{\rm{RTP}}_{t,b_2},\Upsilon^{\rm{RTP}}_{t,b_3}),\\
\max(\Upsilon^{\rm{RP}}_{t,c_1},\Upsilon^{\rm{RP}}_{t,c_2},\Upsilon^{\rm{RP}}_{t,c_3},\Upsilon^{\rm{RP}}_{t,c_4}),\\
\max(\Upsilon^{\rm{RPP}}_{t,d_1},\!\Upsilon^{\rm{RPP}}_{t,d_2},\!\Upsilon^{\rm{RPP}}_{t,d_3},\!\Upsilon^{\rm{RPP}}_{t,d_4},\!\Upsilon^{\rm{RPP}}_{t,d_5}),\\
\end{cases}
\end{align}
respectively. Moreover, $\Upsilon^{sha}_{r,m}$, $\Upsilon^{sha}_{t,m}$, $\Upsilon^{\rm{cyl}}_{r,\rm{upp}}$, and $\Upsilon^{\rm{cyl}}_{t,\rm{upp}}$ can be expressed as
\begin{align} 
&\Upsilon^{sha}_{r,m}=\tan^{-1}\left(\frac{z_{m}\gamma_r}{|x_{m}\cot{\alpha_r}+y_{m}|}\right),\\
&\Upsilon^{sha}_{t,m}=\tan^{-1}\left(\frac{z_{m}\gamma_t}{|x_{m}\cot{\alpha_t}+y_{m}-r|}\right),\label{e14}
\end{align} 
respectively, where $m$ can be $b_1$, $b_2$, $b_3$, $c_1$, $c_2$, $c_3$, $c_4$, $d_1$, $d_2$, $d_3$, $d_4$, and $d_5$. Besides, $\gamma_t$ and $\gamma_r$ can be given by 
\begin{equation}
\gamma_r=\sqrt{\cot^2\alpha_r+1},\,\,\gamma_t=\sqrt{\cot^2\alpha_t+1}.
\end{equation} 

Then, $\Upsilon^{sha}_{r,\min}$ and $\Upsilon^{sha}_{r,\max}$ can be derived as
\begin{align}
\Upsilon^{sha}_{r,\rm{min}}=&
\begin{cases}
\min(\Upsilon^{\rm{RTP}}_{r,b^1_1},\Upsilon^{\rm{RTP}}_{r,b^2_2},\Upsilon^{\rm{RTP}}_{r,b^3_3}),\\
\min(\Upsilon^{\rm{RP}}_{r,c^1_1},\Upsilon^{\rm{RP}}_{r,c^2_2},\Upsilon^{\rm{RP}}_{r,c^3_3},\Upsilon^{\rm{RP}}_{r,c^4_4}),\\
\min(\Upsilon^{\rm{RPP}}_{r,d^1_1},\!\Upsilon^{\rm{RPP}}_{r,d^2_2},\!\Upsilon^{\rm{RPP}}_{r,d^3_3},\!\Upsilon^{\rm{RPP}}_{r,d^4_4},\!\Upsilon^{\rm{RPP}}_{r,d^5_5}),\\
\end{cases}\\
\Upsilon^{sha}_{r,\rm{max}}=&
\begin{cases}
\max(\Upsilon^{\rm{RTP}}_{r,b^1_1},\Upsilon^{\rm{RTP}}_{r,b^2_2},\Upsilon^{\rm{RTP}}_{r,b^3_3}),\\
\max(\Upsilon^{\rm{RP}}_{r,c^1_1},\Upsilon^{\rm{RP}}_{r,c^2_2},\Upsilon^{\rm{RP}}_{r,c^3_3},\Upsilon^{\rm{RP}}_{r,c^4_4}),\\
\max(\!\Upsilon^{\rm{RPP}}_{r,d^1_1},\!\Upsilon^{\rm{RPP}}_{r,d^2_2},\!\Upsilon^{\rm{RPP}}_{r,d^3_3},\!\Upsilon^{\rm{RPP}}_{r,d^4_4},\!\Upsilon^{\rm{RPP}}_{r,d^5_5}),\\
\end{cases}
\end{align} 
respectively, where $\Upsilon^{sha}_{r,o^{i}_{j}}$ denotes the angle between $\boldsymbol{{\rm{RP}}_{r,o^{i}_{j}}}$ and $\boldsymbol{\rm{RG}}$, ${\rm{P}}_{r,o^{i}_{j}}$ represents the intersection point of the plane $\mathcal{E}_{\vartheta}$ with the line ${\rm{O}}_i {\rm{O}}_j$ (e.g. ${\rm{P}}_{r,b^{1}_{1}}$ and line ${\rm{B}}_1 {\rm{B}}'_1$), and RG represents the intersection ray between planes YRZ and $\mathcal{E}_{\vartheta}$. After algebraic operations, $\Upsilon^{sha}_{r,o^{i}_{j}}$ can be expressed as 
\begin{equation}
\Upsilon^{sha}_{r,o^{i}_{j}}=\cos^{-1}\left(\frac{y_{r,o^{i}_{j}}\mathcal{G}_3-z_{r,o^{i}_{j}}\mathcal{G}_2}{\sqrt{\mathcal{G}^2_2+\mathcal{G}^2_3}||\boldsymbol{{\rm{RP}}_{r,o^{i}_{j}}}||}\right),                               
\end{equation} 
where 
\begin{equation}
\begin{cases}
\mathcal{G}_1=-\cos{\alpha_r}\sec^2{\vartheta}\sin(2\phi)\tan{\theta},\\
\mathcal{G}_2=-\sin{\alpha_r}\sec^2{\vartheta}\sin(2\phi)\tan{\theta}, \\
\mathcal{G}_3=2\sec^2{\vartheta}\cos^2{\phi}\tan{\theta},
\end{cases}
\end{equation}
and $\theta$ can be given by $\tan^{-1}(\sec{\phi}\sqrt{\tan^2{\beta_r}-\tan^2{\vartheta}}\cos{\vartheta})$. Besides, $x_{r,o^{i}_{j}}=x_{\xi_{b,c,d}}$, $y_{r,o^{i}_{j}}=y_{\xi_{b,c,d}}$, and $z_{r,o^{i}_{j}}$ can be given by
\begin{equation} 
z_{r,o^{i}_{j}}=-(\mathcal{G}_1 x_{r,o^{i}_{j}}+\mathcal{G}_2\,y_{r,o^{i}_{j}})/\mathcal{G}_3.
\end{equation}  

Following that, $\Upsilon^{sha}_{t,\min}$ and $\Upsilon^{sha}_{t,\max}$ can be derived as
\begin{align}
\Upsilon^{sha}_{t,\rm{min}}=&
\begin{cases}
\min(\Upsilon^{\rm{RTP}}_{t,b^1_1},\Upsilon^{\rm{RTP}}_{t,b^2_2},\Upsilon^{\rm{RTP}}_{t,b^3_3}),\\
\min(\Upsilon^{\rm{RP}}_{t,c^1_1},\Upsilon^{\rm{RP}}_{t,c^2_2},\Upsilon^{\rm{RP}}_{t,c^3_3},\Upsilon^{\rm{RP}}_{t,c^4_4}),\\
\min(\Upsilon^{\rm{RPP}}_{t,d^1_1},\!\Upsilon^{\rm{RPP}}_{t,d^2_2},\!\Upsilon^{\rm{RPP}}_{t,d^3_3},\!\Upsilon^{\rm{RPP}}_{t,d^4_4},\!\Upsilon^{\rm{RPP}}_{t,d^5_5}),\\
\end{cases}\\
\Upsilon^{sha}_{t,\rm{max}}=&
\begin{cases}
\max(\Upsilon^{\rm{RTP}}_{t,b^1_1},\Upsilon^{\rm{RTP}}_{t,b^2_2},\Upsilon^{\rm{RTP}}_{t,b^3_3}),\\
\max(\Upsilon^{\rm{RP}}_{t,c^1_1},\Upsilon^{\rm{RP}}_{t,c^2_2},\Upsilon^{\rm{RP}}_{t,c^3_3},\Upsilon^{\rm{RP}}_{t,c^4_4}),\\
\max(\Upsilon^{\rm{RPP}}_{t,d^1_1},\!\Upsilon^{\rm{RPP}}_{t,d^2_2},\!\Upsilon^{\rm{RPP}}_{t,d^3_3},\!\Upsilon^{\rm{RPP}}_{t,d^4_4},\!\Upsilon^{\rm{RPP}}_{t,d^5_5}),\\
\end{cases}
\end{align}   
respectively, where $\Upsilon^{sha}_{t,o^{i}_{j}}$ represents the angle between $\boldsymbol{{\rm{TP}}_{t,o^{i}_{j}}}$ and $\boldsymbol{\rm{TH}}$, ${\rm{P}}_{t,o^{i}_{j}}$ denotes the intersection point of the plane $\mathcal{F}_{\sigma}$ with the line ${\rm{O}}_i {\rm{O}}_j$ (e.g. ${\rm{P}}_{t,b^{1}_{1}}$ and line ${\rm{B}}_1 {\rm{B}}'_1$), and TH represents the intersection ray between planes YRZ and $\mathcal{F}_{\sigma}$. Here, $\mathcal{F}_{\sigma}$ is defined as the plane that is perpendicular to the plane TVV' and rotates around the line TL, where TL is located on the XRY and perpendicular to the line TV', and the subscript $\sigma$ denotes the angle between the plane $\mathcal{F}_{\sigma}$ and the plane LTV', which is positive if rotating clockwise from LTV'. After algebraic operations, $\Upsilon^{sha}_{t,o^{i}_{j}}$ can be expressed as
\begin{equation}  
\Upsilon^{sha}_{t,o^{i}_{j}}=\cos^{-1}\left[\frac{(r-y_{t,o^{i}_{j}})\cos{\sigma}-z_{t,o^{i}_{j}}\sin{\alpha_t}\sin{\sigma}}{\sqrt{\sin^2{\alpha_t}\sin^2{\sigma}+\cos^2{\sigma}}||\boldsymbol{{\rm{TP}}_{t,o^{i}_{j}}}||}\right],
\end{equation}
where $x_{t,o^{i}_{j}}=x_{\xi_{b,c,d}}$, $y_{t,o^{i}_{j}}=y_{\xi_{b,c,d}}$, and $z_{t,o^{i}_{j}}$ can be given by   
\begin{equation}
z_{t,o^{i}_{j}}={\sin{\sigma}\cos{\alpha_t}\,x_{t,o^{i}_{j}}+\sin{\sigma}\sin{\alpha_t} (y_{t,o^{i}_{j}}-r)}/{\cos{\sigma}}.                               
\end{equation}  

Moreover, we investigate the intersection cases between the receiver FoV and the obstacle, where the boundary parameters $\Psi^r_{\min}$ and $\Psi^r_{\max}$ of the receiver FoV can be derived as  
\begin{equation}
\Psi^r_{\min}\,=\alpha_r+\psi_{\min}-\frac{\pi}{2}, \Psi^r_{\max}=\alpha_r+\psi_{\max}-\frac{\pi}{2}.
\end{equation} 
For tractable analysis, we suppose that $\Upsilon^{sha}_{r,\min}$, $\Upsilon^{sha}_{r,\max}$, $\Psi^r_{\min}$, and $\Psi^r_{\max}$ satisfy the relation that $\Psi^r_{\max}>\Upsilon^{sha}_{r,\max}>\Upsilon^{sha}_{r,\min}>\Psi^r_{\min}$, since this assumption contains the most comprehensive possibilities. The derivation of $\mathcal{S}_{\rm{wei}}$ is divided into the following three steps:

\textbf{Step 1:} Judge whether the coordinates $(x,y,z)$ determined by (\ref{e6}), (\ref{e7}), (\ref{e8}), and (\ref{e10}) satisfy inequality (\ref{e9}). If not, assign a value of 0 to $\mathcal{S}_{\rm{wei}}$.  

\textbf{Step 2:} For the situation where $\phi\ge\Upsilon^{sha}_{r,\rm{upp}}$ and $\sigma\ge\Upsilon^{sha}_{t,\rm{upp}}$, $\mathcal{S}_{\rm{wei}}=1$, where $\sigma$ can be given by $\varphi$, ${\pi}/{2}$, and $\pi-\varphi$ for $y+x\cot{\alpha_t}$ less than $r$, equal to $r$, and greater than $r$, respectively, and $\varphi$ can be determined through (\ref{e14}). For the situation where $\phi\ge\Upsilon^{sha}_{r,\rm{upp}}$ and $\sigma<\Upsilon^{sha}_{t,\rm{upp}}$, the value of $\mathcal{S}_{\rm{wei}}$ equal to 1 has two circumstances: i) $\Psi^t_{\rm{esp}}$ is less than $\Upsilon^{sha}_{t,\rm{min}}$ or greater than $\Upsilon^{sha}_{t,\rm{max}}$, where $\Psi^t_{\rm{esp}}$ is the angle between $\boldsymbol{{\rm{TP}}}$ and $\boldsymbol{\rm{TH}}$, and ii) $\Lambda^t_{\rm{tmp}}<||\boldsymbol{{\rm{TO}}'_\xi}||$, $\Upsilon^{sha}_{t,\rm{min}}\le\Psi^t_{\rm{esp}}\le\Upsilon^{sha}_{t,\rm{max}}$, and $\Gamma_{\rm{tmp}}>\Gamma^{sha}_{\rm{rad}}$, where $\Gamma_{\rm{tmp}}$ is the distance from the scattering point $\rm{P}$ to the central axis of the obstacle, and $\Lambda^t_{\rm{tmp}}$ is the horizontal distance from $\rm{P}$ to $\rm{T}$.

\textbf{Step 3:} For the situation where $\phi<\Upsilon^{sha}_{r,\rm{upp}}$, if $\Psi^r_{\rm{esp}}<\Upsilon^{sha}_{r,\rm{min}}$, the value of $\mathcal{S}_{\rm{wei}}$ is equal to 1, where $\Psi^r_{\rm{esp}}$ is the angle between $\boldsymbol{{\rm{RP}}}$ and $\boldsymbol{\rm{RG}}$, while if $\Psi^r_{\rm{esp}}\ge\Upsilon^{sha}_{r,\rm{min}}$, the value of $\mathcal{S}_{\rm{wei}}$ equal to 1 has two circumstances: i) $\sigma\ge\Upsilon^{sha}_{t,\rm{upp}}$, and ii) $\sigma<\Upsilon^{sha}_{t,\rm{upp}}$ and $\Psi^t_{\rm{esp}}<\Upsilon^{sha}_{t,\rm{min}}$. Here, two constraints need to be imposed on the scattering points for $\Upsilon^{sha}_{r,\rm{min}}\le\Psi^r_{\rm{esp}}\le\Upsilon^{sha}_{r,\rm{max}}$: 1) $\Gamma_{\rm{tmp}}>\Gamma^{sha}_{\rm{rad}}$, and 2) $\Lambda^r_{\rm{tmp}}<||\boldsymbol{{\rm{RO}}'_\xi}||$, where $\Lambda^r_{\rm{tmp}}$ denotes the horizontal distance from $\rm{P}$ to $\rm{R}$. For $\Psi^r_{\rm{esp}}>\Upsilon^{sha}_{r,\rm{max}}$ and ii) $\sigma<\Upsilon^{sha}_{t,\rm{upp}}$, the situation where $\Psi^t_{\rm{esp}}>\Upsilon^{sha}_{t,\rm{max}}$ also needs to be incorporated.

\section{Reflected Energy Derivation}
In this part, we derive the received pulse energy contributed by obstacle reflection. Based on the existing reflection propagation theory \cite{ref7}, $\mathcal{Q}_{r,\rm{ref}}$ can be derived as
\begin{equation}
\begin{aligned}
\mathcal{Q}_{r,\rm{ref}}=&\iint_{\mathcal{S}} \frac{(\kappa+1)\cos^{\kappa}\varpi\cos{\vartheta_i}\cos{\delta}}{2\pi\varepsilon^2\nu^2}\\
                                       & \times \mathcal{Q}_t \mathcal{A}\,v_r\,\Delta(\vartheta_1,\vartheta_2)\exp[-k_e(\varepsilon+\nu)]\,{\rm{d}}s,
\end{aligned}
\label{e26}
\end{equation} 
where $\vartheta_i$ is the incidence angle between the normal vector $\boldsymbol{n}$ of the reflection region and $-\boldsymbol{\varepsilon}$; $v_r$ is the reflection coefficient; $\vartheta_1$ and $\vartheta_2$ represent the angles between $\boldsymbol{n}$ and $\boldsymbol{\nu}$, and $\boldsymbol{\nu}$ and $\boldsymbol{\varepsilon_s}$, respectively. Besides, $\boldsymbol{\varepsilon_s}$ is the direction vector of the specular reflection of $\boldsymbol{\varepsilon}$ and $\Delta(\vartheta_1,\vartheta_2)$ denotes the reflection pattern of the obstacle, which can be expressed as \cite{ref14} 
\begin{equation}  
\Delta(\vartheta_1,\vartheta_2)=\mu\frac{\cos{\vartheta_1}}{\pi}+(1-\mu)\frac{m_s+1}{2\pi}\cos^{m_s}\vartheta_2,   
\end{equation}
where $\mu$ is the percentage of the incident signal that is reflected diffusely and supposes values between $0$ and $1$, and $m_s$ is the directivity of the specular components. The effective reflection region $\mathcal{S}$ and differential reflection area d$s$ will be determined combined with specific obstacle orientation angles.

Combined with the geometries of RTP, RP, and RPP, they all have three potential reflection surfaces, which can be given by
\begin{equation}
\mathcal{S}_{sha}:
\begin{cases}
y\in[y_{\varrho_i},y_{\varrho_j}],&{\rm{surface}}\,\,\mathbb{W}_{i j j' i'},\\
x=k_{\varrho_{ij}}+x_{\varrho_j},&{\rm{surface}}\,\,\mathbb{W}_{i j j' i'},\\
z\in[0,\gamma_\varrho],&{sha},
\end{cases}\label{e28}\\
\end{equation}
where when $sha$ denotes RTP, $\varrho$ and $\mathbb{W}$ are $b$ and $\rm{B}$, respectively, $(i,j)$ can be $(1,3)$, $(3,2)$, and $(2,1)$, and $\mathbb{W}_{i j j' i'}$ denotes the surface ${\rm{B}}_i {\rm{B}}_j {\rm{B}}'_j {\rm{B}}'_i$; when $sha$ denotes RP, $\varrho$ and $\mathbb{W}$ are $c$ and $\rm{C}$, respectively, and $(i,j)$ can be $(4,3)$, $(3,2)$, and $(2,1)$; when $sha$ denotes RPP, $\varrho$ and $\mathbb{W}$ are $d$ and $\rm{D}$, respectively, $(i,j)$ can be $(5,4)$, $(4,3)$, and $(3,2)$. Besides, the $\alpha_b$ intervals for $\rm{B}_{1 3 3' 1'}$, $\rm{B}_{3 2 2' 3'}$, and $\rm{B}_{2 1 1' 2'}$ are $[0,\pi/6)$, $[0,2\pi/3]$, and $(\pi/2,2\pi/3]$, respectively; the $\alpha_c$ intervals for $\rm{C}_{4 3 3' 4'}$, $\rm{C}_{3 2 2' 3'}$, and $\rm{C}_{2 1 1' 2'}$ are $[0,\pi/2)$, $(0,\pi)$, and $(\pi/2,\pi]$, respectively; and $\alpha_d$ intervals for $\rm{D}_{5 4 4' 5'}$, $\rm{D}_{4 3 3' 4'}$, and $\rm{D}_{3 2 2' 3'}$ are $[0,3\pi/10)$, $[0,2\pi/5]$, and $(\pi/10,2\pi/5]$, respectively. On these bases, $k_{\varrho_{ij}}$ can be derived as
\begin{equation}
k_{\varrho_{ij}}=\frac{(x_{\varrho_i}-x_{\varrho_j})(y-y_{\varrho_j})}{y_{\varrho_i}-y_{\varrho_j}},   
\end{equation}
and the normal vector of these surfaces can be given by 
\begin{equation}
\boldsymbol{n_{\varrho_i}}=[\gamma_\varrho(y_{\varrho_j}-y_{\varrho_i}),\gamma_\varrho(x_{\varrho_i}-x_{\varrho_j}),0].
\end{equation} 
Further, the coordinates $(x,y,z)$ must satisfy the relationships that ${\boldsymbol{\mu_t}}{\boldsymbol{\varepsilon}}^{\rm{T}}\geq0$ and ${\boldsymbol{\mu_r}}(-{\boldsymbol{\nu}})^{\rm{T}}\ge\nu\cos{\beta_r}$, and the expressions of ${\rm{d}}s$ can be derived as
\begin{equation}
{\rm{d}}s=
\begin{cases}
\frac{{\rm{d}}y{\rm{d}}z}{\sin(\pi/6-\alpha_b)},&\rm{surface}\,\,B_{1 3 3' 1'},\\[4pt]
\frac{{\rm{d}}y{\rm{d}}z}{\sin(\pi/6+\alpha_b)},&\rm{surface}\,\,B_{3 2 2' 3'},\\[4pt]
\frac{{\rm{d}}y{\rm{d}}z}{\sin(\alpha_b-\pi/2)},&\rm{surface}\,\,B_{2 1 1' 2'},\\[4pt]
\frac{{\rm{d}}y{\rm{d}}z}{\cos\alpha_c},&\rm{surface}\,\,C_{4 3 3' 4'},\\[4pt]
\frac{{\rm{d}}y{\rm{d}}z}{\sin\alpha_c},&\rm{surface}\,\,C_{3 2 2' 3'},\\[4pt]
\frac{{\rm{d}}y{\rm{d}}z}{-\cos\alpha_c},&\rm{surface}\,\,C_{2 1 1' 2'},\\[4pt]
\frac{{\rm{d}}y{\rm{d}}z}{\sin(3\pi/10-\alpha_d)},&\rm{surface}\,\,D_{5 4 4' 5'},\\[4pt]
\frac{{\rm{d}}y{\rm{d}}z}{\cos(\alpha_d-\pi/5)},&\rm{surface}\,\,D_{4 3 3' 4'},\\[4pt]
\frac{{\rm{d}}y{\rm{d}}z}{\sin(\alpha_d-\pi/10)},&\rm{surface}\,\,D_{3 2 2' 3'}.
\end{cases}
\end{equation}   

Note that in the derivation process of $\mathcal{S}$, we defaulted to the fact that the angle between $\boldsymbol{n_{\varrho_i}}$ and $-{\boldsymbol{\varepsilon}}$ as well as the angle between $\boldsymbol{n_{\varrho_i}}$ and ${\boldsymbol{\nu}}$ are both less than $\pi/2$.  

\section{Numerical Results}  
\begin{figure}[t]  
\centering  
\includegraphics[scale=0.395]{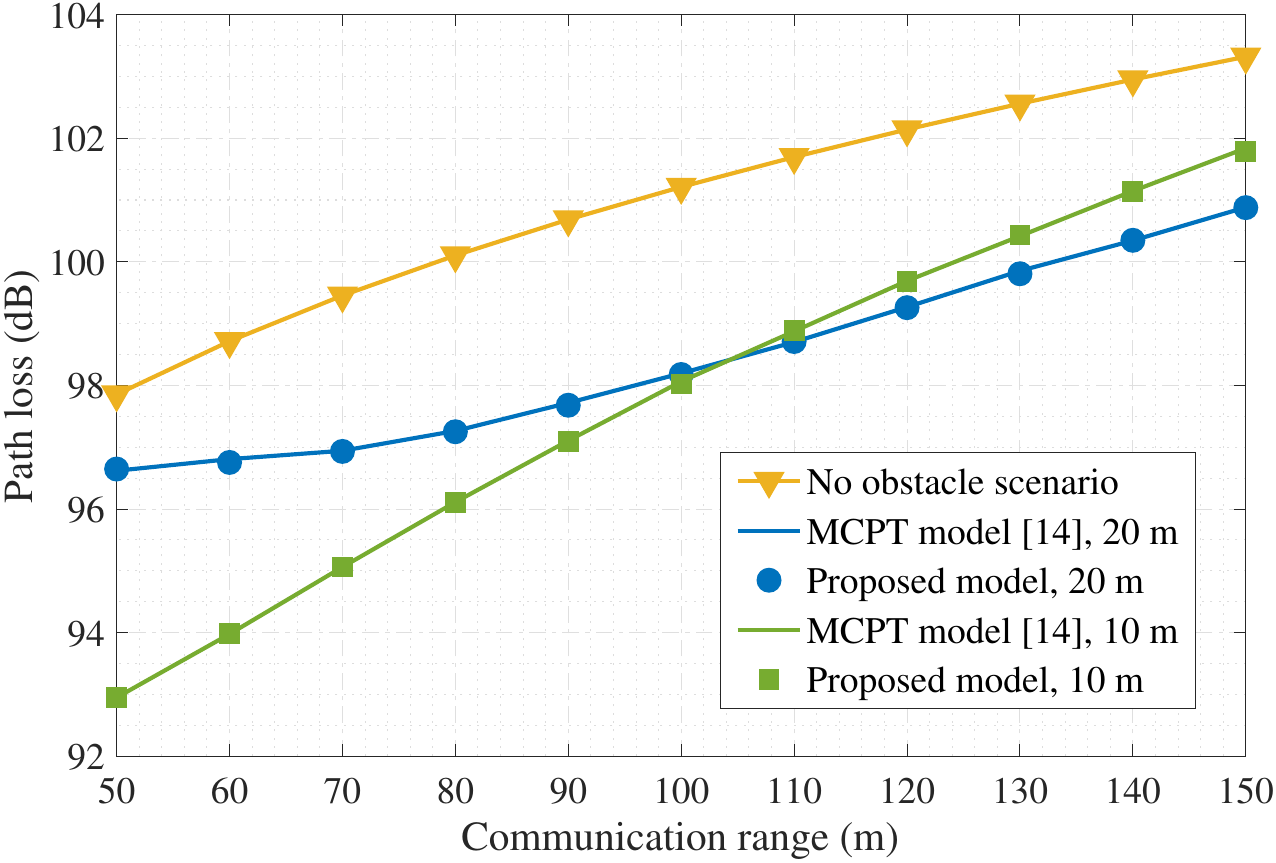}
\centering
\caption{Path loss results for the proposed model and the MCPT model under different scenarios.}
\label{Fig4}  
\end{figure}

In this section, we examine the performance of the proposed path loss model and analyze the impacts of different obstacle contours on the path loss of NLoS UV channels. As shown in Fig.~\ref{Fig4}, the proposed model is verified against the MCPT benchmark \cite{ref14} with an RP obstacle. Key parameters align with \cite{ref14}: $\mu_c=r/10$, $\omega_c=3r$, $\gamma_c=3r$, $x_c=-\mu_c/2-10\,\rm{m} (\rm{or}\,-20\,\rm{m})$, $y_c=r/2$, $\alpha_c=0^{\circ}$, $v_r=0.1$, $\mu=0.5$, $m_s=5$, and $A_r=1.92\,\rm{cm}^2$. Simulations employ $10^{7}$ photons with a survival probability threshold of $10^{-10}$, while remaining parameters are detailed in Table~\ref{t1}. The numerical results demonstrate strong agreement between the path loss curves of the proposed model and the MCPT benchmark across scenarios with reflection surface-to-transceiver distances of $10\,\rm{m}$ and $20\,\rm{m}$. Notably, the NLoS channel without obstacles exhibits higher path loss than its obstructed counterpart, attributed to the significant energy contribution from obstacle reflections.   

\linespread{1.5}
\begin{table}[t]
\centering
\linespread{1.0}
\caption{Parameter Settings for the Proposed Model and MCPT Model}
\label{t1}
\begin{tabular}{l l || l l}
\hline
\hline
{Parameters}&{Values}&{Parameters}&{Values}\\
\hline
\hline
$k^{\rm{Ray}}_s$&$0.24\,\rm{km^{-1}}$&$\beta_{1/2}$&$\pi/6$\\
$k^{\rm{Mie}}_s$&$0.25\,\rm{km^{-1}}$&$\beta_r$&$\pi/6$\\
$k_a$&$0.90\,\rm{km^{-1}}$&$\alpha_t$&$-19\pi/36$\\
$\gamma$&$0.017$&$\alpha_r$&$19\pi/36$\\
g&0.72&$\vartheta_t$&$7\pi/36$\\
$f$&0.5&$\vartheta_r$&$7\pi/36$\\
\hline
\hline
\end{tabular}
\end{table}
\linespread{1.0} 
\begin{figure}[t]  
\centering  
\includegraphics[scale=0.395]{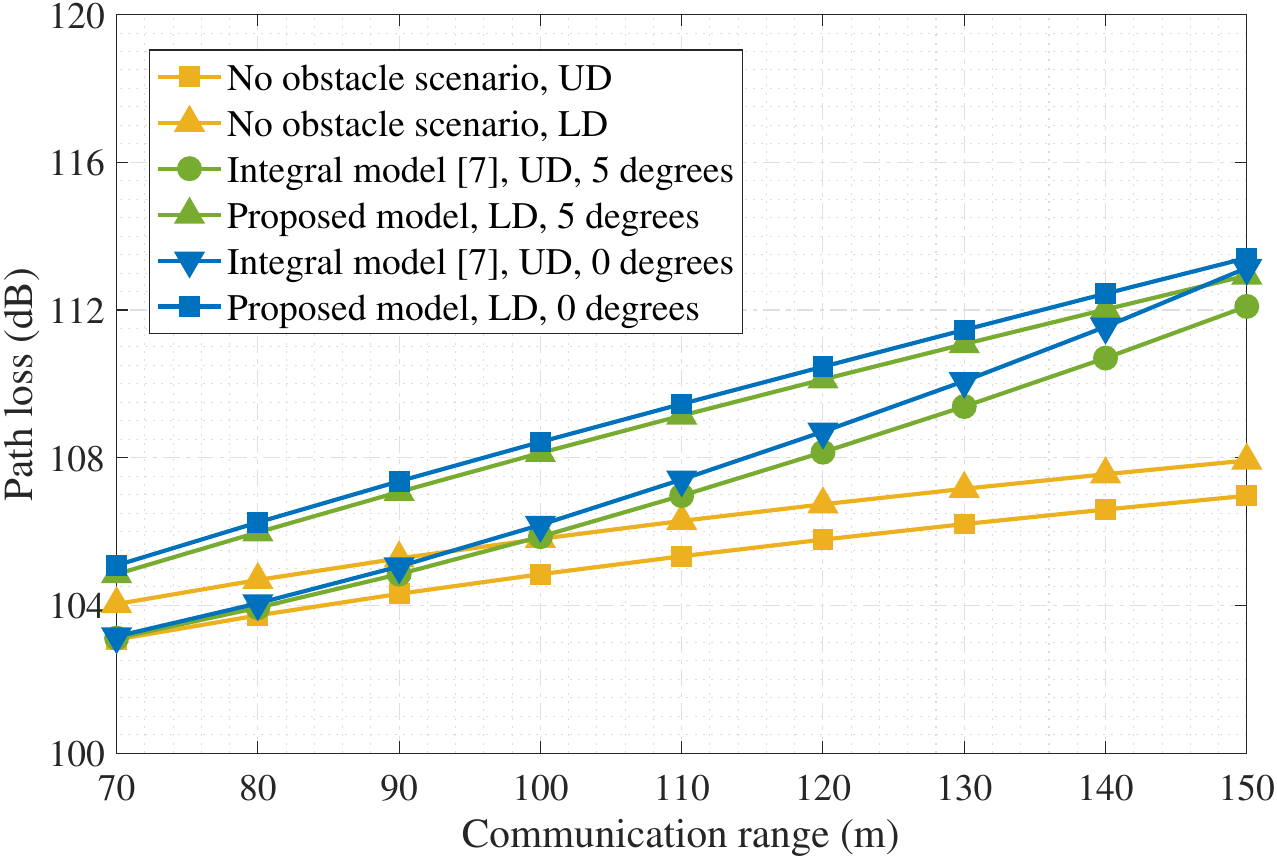}
\centering
\caption{Path loss results of the scattered energy for the proposed model and integral model under different obstacle orientation angles.}
\label{Fig5}  
\end{figure} 
\linespread{1.5}
\begin{table}[t]
\centering
\linespread{1.0}
\caption{Parameter Settings for the Transceiver and Obstacle}
\label{t2}
\begin{tabular}{l l || l l}
\hline
\hline
{Parameters}&{Values}&{Parameters}&{Values}\\
\hline
\hline
$\beta_{1/2}$&$\pi/12$&$\omega_c$&40 m\\
$\beta_r$&$\pi/12$&$\gamma_c$&80 m\\
$\alpha_t$&$-2\pi/3$&$\mu_c$&30 m\\
$\alpha_r$&$2\pi/3$&$\alpha_c$&$0^{\circ},\,5^{\circ}$\\
$\vartheta_t$&$\pi/9$&$\vartheta_r$&$\pi/9$\\
\hline
\hline
\end{tabular}
\end{table}
\linespread{1.0} 

As shown in Fig.~\ref{Fig5}, the influences of the proposed obstacle-boundary approximation method on the path loss estimation are investigated based on the latest work \cite{ref7}, where the contour of the obstacle is chosen as an RP. The parameter settings of the transceiver and obstacle are provided in Table~\ref{t2}, where $x_c$ and $y_c$ can be expressed as $-\Gamma^{\rm{RP}}_{\rm{rad}}\sin[\tan^{-1}(\mu_c/\omega_c)+\alpha_c]-\mu_c$ and $r/2$, respectively, and the rest parameters are consistent with Table~\ref{t1}. UD/LD denote uniform/Lambertian radiation distributions. Fig.~\ref{Fig5} shows that without obstacles, path loss differences between UD and LD remain about 0.95 dB across all tested distances. On this basis, the maximum error caused by boundary simplification stays below 1.35 dB for $\alpha_c=0^{\circ}$ and 1.30 dB for $5^{\circ}$, which is negligible when obstacle reflection dominates signal propagation.     

\begin{figure}[t]  
\centering  
\includegraphics[scale=0.395]{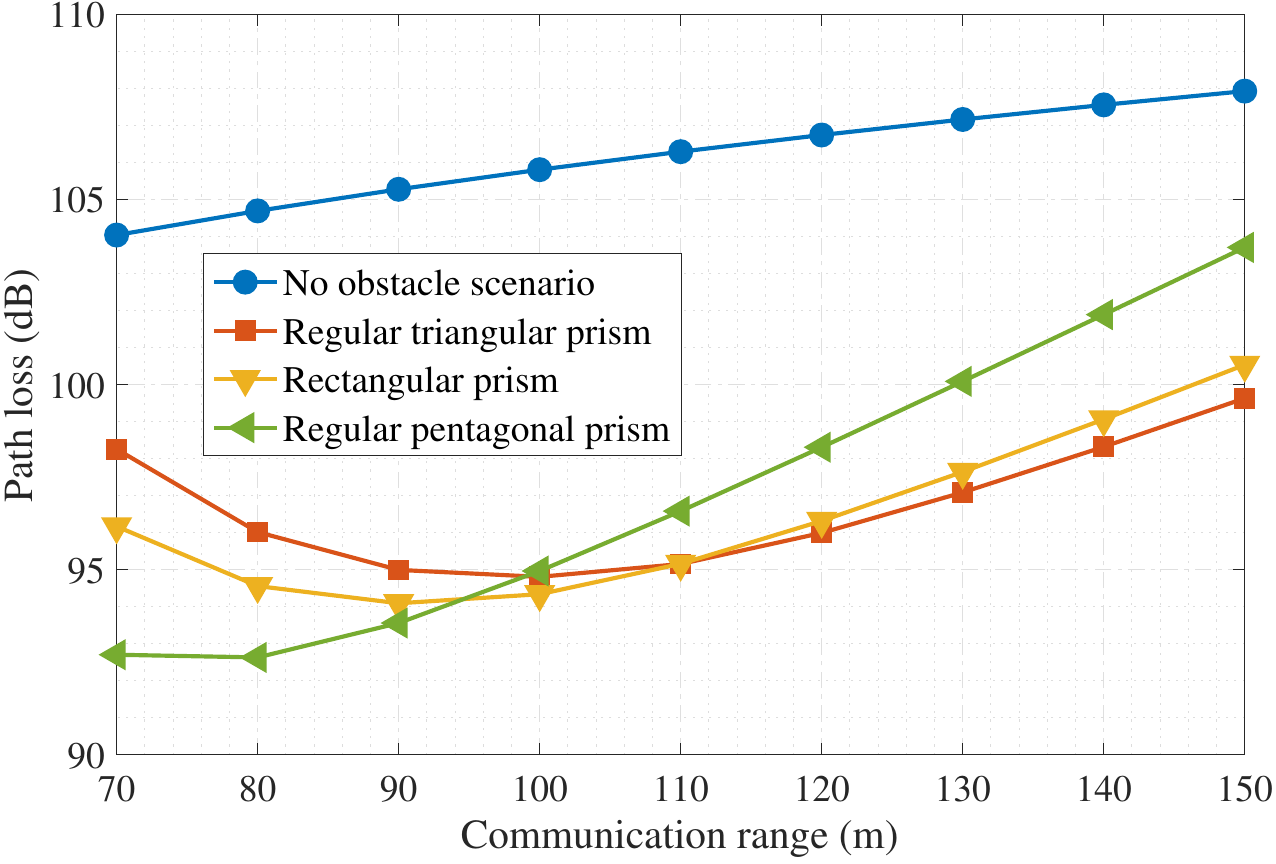}
\centering
\caption{Path loss results for NLoS UV communication channels incorporating different obstacle contours.}
\label{Fig6}  
\end{figure}

Moreover, we investigate the impacts of the obstacle contours on the channel path loss under identical transceiver parameter settings, as shown in Fig.~\ref{Fig6}. The system model parameters are selected as follows: $x_\xi$, $y_\xi$, $z_\xi$, and $\Gamma^{sha}_{\rm{rad}}$ are set to $-46.7\,\rm{m}$, $r/2$, $80\,\rm{m}$, and $25\,\rm{m}$, respectively, and $\alpha_a$, $\alpha_b$, $\alpha_c$, and $\alpha_d$ are selected as $\pi/3$, $\pi/3$, $0$, and $\pi/5$, respectively. As for the rest parameter settings, they are in accordance with Table~\ref{t2}. From Fig.~\ref{Fig6} we can discover that among the three obstacle contours, the RPP contributes most to the received energy at short communication ranges, and the RTP dominates the received energy at long communication ranges.

\section{Conclusion}
This paper addressed critical limitations in prior NLoS UV channel modeling by developing a comprehensive path loss model incorporating scattering and obstacle interactions. Departing from conventional obstacle-free models relying on uniform intensity assumptions, our approach integrated Lambertian radiation patterns and introduced an efficient obstacle-boundary approximation method to address complex spatial characteristics. Numerical validation through MCPT and analytical benchmarking demonstrated that boundary simplification induced negligible attenuation estimation errors under dominant reflection conditions. This work provided crucial guidelines for the design of NLoS UV communication systems. 
   
\section*{Acknowledgement}  
This work was supported by the National Key Research and Development Program of China (2023YFE0110600), was partially supported by the Open Fund Project of Hanjiang National Laboratory (No.KF2024027), was partially supported by the National Natural Science Foundation of China (62401433), and was partially supported by NSF ECCS-2302469, Toyota. Amazon and Japan Science and Technology Agency (JST) Adopting Sustainable Partnerships for Innovative Research Ecosystem (ASPIRE) JPMJAP2326. 

\end{document}